\documentclass[final,3p,times,twocolumn]{elsarticle}

\usepackage{amssymb}

\journal{Physics Letters B}

\def \b{\beta}

\def \be{\begin{equation}}
\def \ee{\end{equation}}
\def \ben{\begin{eqnarray}}
\def \een{\end{eqnarray}}

\def \O{\Omega}
\def \L{\Lambda} 
 
\begin{document}

\begin{frontmatter}

\title{Dark energy from quantum wave function collapse of dark matter}
 
\author{A. S. Majumdar}
 
\address{S. N. Bose National Centre for Basic Sciences,
Block JD, Sector III, Salt Lake, Kolkata 700098, India}

\ead{archan@bose.res.in}

\author{D. Home}

\address{Bose Institute, Kolkata 700009, India}

\ead{dhome@bosemain.boseinst.ac.in}

\author{S. Sinha}

\address{Raman Research Institute, Sadashivanagar, Bangalore 560080}

\ead{meetsiddhartha@gmail.com}

\begin{abstract} 
 
Dynamical wave function collapse models entail the continuous liberation of
a specified rate of energy arising from the interaction of a fluctuating 
scalar field with the matter wave function. We consider 
the wave function collapse process for the constituents of dark matter in
our universe. Beginning from a particular early era of the universe chosen from
physical considerations, the rate of the associated energy 
liberation is integrated to yield the requisite magnitude of dark energy
around the era of galaxy formation.
Further, the equation of state for the liberated energy approaches $w \to -1$
asymptotically, providing a mechanism to generate the present acceleration of 
the universe.
\end{abstract}


\end{frontmatter}

\section{Introduction}

More than a century of development of physical theory since the advent of 
quantum mechanics and relativity has led to profound advancements of our 
understanding of the microcosm
as well as the macrocosm. Yet certain deep mysteries have emerged in the
study of particular aspects like the quantum measurement 
problem\cite{measure} of the former, and the mechanism for the presently
accelerating universe\cite{accel} of the latter arena. Satisfactory 
resolution of these two fundamental challenges encountered by modern 
physics may call for 
close introspection of any possible domain of overlap between the solutions
that have been offered separately for either of them. 

The linearity and unitary evolution of
quantum theory give rise to the entanglement of elementary
particles having widespread and fascinating 
applications\cite{entang}. Basic quantum theory predicts the 
persistence of quantum entanglement even for macrosystems\cite{large}.
However, in practice, it is difficult to realize the quantum entanglement
of macroscopic entities over
large distance and time scales. The emergence
of classicality observed in the real world is hard to understand in terms
of any simple limiting behaviour of quantum theory. A key issue not explained
by quantum theory is {\it how} a definite outcome occurs as the result
of an individual measurement on a quantum system\cite{measure}. Over the 
years several approaches have been suggested to tackle this problem,
such as environment induced 
decoherence\cite{decoh}, the quantum state diffusion picture\cite{qsd},
the consistent
histories approach\cite{consistent}, the Bohmian ontological 
interpretation\cite{bohm},  
and the dynamical models of wave function
collapse\cite{csl,masscsl,gravcoll,exptcsl,rcsl}.

In dynamical collapse models wave function collapse is regarded as a real
physical process describing the measurement 
dynamics without discontinuity or added interpretations.
In the `Spontaneous localization' models\cite{csl} 
the unitary Schrodinger evolution is modified by stochastic nonlinear
terms that affect the dynamics on time scales relevant to typical 
macrophysical situations. The emergence of classicality and the ocurrence
of single outcomes in measurements is achieved by the interaction of the
fluctuating modes of a scalar field with the relevant wave functions at a 
rate proportional to the number or the mass\cite{masscsl} of the particles 
involved. The mass-dependent collapse rate is somewhat similar to the spirit 
of gravity induced state vector reduction\cite{gravcoll}. 
Recently, relativistic generalizations of collapse models
have been made\cite{rcsl}, displaying the conservation of the energy 
exchanged between the scalar field and
the collapsing matter.
In essence, the dynamical collapse models are able to achieve the 
quantum to classical transition within standard quantum theory with just
the additional postulated role played by a fluctuating scalar field. It
is thus desirable that the existence of such an energy liberating
scalar field be motivated by some other physical considerations.

In the cosmological arena one or more scalar
fields have been invoked to acount for the observed features of the 
universe since 
its very early stages. The inflationary
paradigm  based on the dominant scalar field energy is widely
accepted as an essential extension
of the standard big-bang cosmological 
model\cite{books}.
Scalar fields play central roles in unified particle physics 
(electroweak and grand-unification) models and string- and brane-theory 
models as well, and much of the physics of the early universe is inspired
by these models\cite{books}. 

Observations of high redshift Type Ia Supernovae (SN Ia) 
 \cite{accel} led to the conclusion that our universe is presently undergoing
a phase of accelerated expansion. This behaviour of the present universe
is possible through the presence of a dominant ``dark'' energy component.
Apart from SN Ia observations, indirect evidences from CMB anisotropy
 and large-scale structure studies show that the dark energy constitutes 
about $70$\% of the 
total energy density of the universe at present \cite{cmbr-lss}, is smoothly 
distributed in 
space, and has large negative pressure. Several possibilites, such as the 
existence of a cosmological constant, for dark-energy 
candidates have been proposed, 
(see Ref.~\cite{DErev} for reviews). The idea that a 
scalar field rolling
along the slope of its potential (quintessence\cite{quint} or 
{\it k}-essence\cite{kessence} models) provides the 
required amount 
of dark energy has gained some popularity.  A generic
feature of such models is the ``ad-hoc'' construction of the scalar field
potentials to ensure compatibility with 
observational constraints\cite{DErev}. The problem of 
``cosmic coincidence'' \cite{DErev},  as to why
the scalar field energy density starts dominating just before
the present era, remains. 

The motivation for this paper is to look for
a possible connection between these two independently well-founded
proposals involving scalar fields in separate domains, namely dynamical 
wave function
collapse advocated for
the emergence of classicality of the quantum world, and the mechanism for
the scalar field driven present acceleration of the universe, respectively.
Our purpose here 
is to gain additional insights on these problems in a scenario where
the cosmic scalar field causes dynamical wave function collapse
of the constituents of dark matter in the universe.

\section{The scheme}

We begin by considering the wave function collapse process in spontaneous 
localization models \cite{csl,masscsl,exptcsl,rcsl} which is triggered for 
a wave function involving one or more particle(s) 
when the scale of their superposition in position space exceeds the value
given by a parameter $a_{*}$.  The associated
rate of energy liberation in mass-dependent
dynamical collapse models\cite{csl,masscsl,rcsl} for a system with mass 
$M$ is given by
\be 
\frac{dE}{dt} = \frac{3\hbar^2 M}{4m_0^2 a_{*}^2T_{*}} \equiv \b
\label{enrate}
\ee
where $m_0 = 10^{-24}{\mathrm g}$,  $a_{*} = 10^{-5}{\mathrm cm}$
and $T_{*} = 10^{16}{\mathrm s}$ are the parameters of the collapse
models\cite{csl,masscsl,exptcsl,rcsl}.  These values are chosen such that
any superposition of the wave functions of microparticles 
like the proton is kept intact over the relevant
distance and time scales, and are consistent with
the results of all laboratory experiments performed so far\cite{exptcsl}. 
Let us consider 
a situation
where there is a uniform distribution of mass in a region
of volume $V$. The rate of energy liberation Eq.(\ref{enrate})
can be written as
\be
\frac{d}{dt}(\rho V) = \b
\label{densrate}
\ee
where $\rho$ is the density of the energy gained by the region $V$. 
If the fluctuating scalar field $\phi$ that drives the collapse has an energy 
density
$\rho_{\phi}$, at any given time a part of it is pumped into the region $V$ 
with an instantaneous rate 
$\dot{\rho}_{\phi}$ such that the rate of energy loss by the field in the
volume $V$ at this particular instance of time is given 
by $\dot{\rho}_{\phi}V$.  Conservation
of energy\cite{rcsl} between the scalar field and the collapsing matter in the
region $V$ dictates that
\be
\dot{\rho}_{\phi}V = -\b
\label{conserv}
\ee
using Eq.(\ref{enrate}). 

We now apply the above arguments to the dynamical collapse of the dark
matter constituents of our universe, driven by the interaction with the
fluctuating modes of a cosmological scalar field. We
have in mind the typical scenario of the early universe where matter is
distributed uniformly in our expanding Robertson-Walker(RW) Hubble volume
with scale factor $R$. 
The expansion of the universe aids the collapse process since the physical
separation between two comoving and entangled wave packets (the physical length
scale of the superpositions in position space) increases with time.
The rate of wave function collapse should be higher at earlier times
since the corresponding rate of expansion is also higher. These considerations
were used to evaluate the effect of the energy liberation due
to the dynamical collapse of baryonic wave functions\cite{pla}. In the
present analysis we focus on the collapse of the dark matter 
since its contribution
to the total energy density of the universe exceeds that of baryonic matter 
by more than
one order of magnitude\cite{cmbr-lss}.
Replacing $\rho$ by $\rho_m$ (where $\rho_m$ is the energy density of 
dark matter), in  Eq(\ref{densrate}), 
one obtains
\be
\dot{\rho}_m + 3\rho_m \frac{\dot{R}}{R} = \frac{\b}{R^3}
\label{darkm1}
\ee
It is apparent that due to the expansion of the universe  all of
the energy supplied by the scalar field (r.h.s. of Eq.(\ref{darkm1})) does
not contribute towards increasing the matter energy.

The activation of the process of dynamical 
collapse of the constituents of dark matter require
the following criteria to hold.  The physical size of the superposed
wave functions  
of the dark matter particles in position space should be comparable to 
or greater than the 
parameter $a_{*}$. Secondly, the
scalar field should possess a finite energy density to drive the collapse.
The integration of Eq.(\ref{darkm1}) over a time 
period during which dynamical collpase of matter in the expanding RW background
is effective with the rate given by Eq.(\ref{darkm1} (say, 
from $t_i$ to $t_{\mathrm end}$) gives the total 
energy density of dark matter at the time $t_{\mathrm end}$, i.e.,
\be
\int_{t_i}^{t_{\mathrm end}} \dot{\rho}_m dt = (\rho_m)_{\mathrm end} 
+ (\rho_m)_{\mathrm extra}
\label{darkm2}
\ee
where $(\rho_m)_{\mathrm end}$ is the standard amount of the dark matter
energy density at time $t_{\mathrm end}$ (that can be obtained from the 
initial density
at time $t_i$ by the usual scaling due to expansion), and 
$(\rho_m)_{\mathrm extra}$ is the excess contribution coming from the 
dynamical collapse process. 

In the standard model of particle physics, mass of particles is created
by the electroweak symmetry breaking. Further, there may not be any matter
(even of exotic types) in the universe before the electroweak phase 
transition, since the matter-antimatter asymmetry in the universe 
may itself be created at that epoch \cite{books}. Any particle created 
earlier is very quickly anhiliated by
its anti-particle. Since we are considering a mass-dependent wave function
collapse scheme in the present analysis, it is justifiable to put a
lower limit on the time scale on which collapse is activated in the
early universe to be of the order of $t_{\mathrm EW}$, the epoch of the
electroweak phase transition in the early universe. Moreover, if
supersymmetry is unbroken, a finite vacuum energy for any field is
not possible (any contribution from a bosonic field gets cancelled by
its fermionic superpartner). Therefore, the vaccum energy of the scalar
field does not exist and thus could not be responsible for driving
collapse before supersymmetry breaking in the early universe. Since,
in several plausible models of high energy physics, supersymmetry is
broken just prior to the epoch of the electroweak phase 
transition \cite{books}, this
further motivates our choice of $t_i = t_{\mathrm EW}$.

On the other hand, 
the rate of energy gain in Eq.(\ref{darkm1}) is not valid beyond
the time when most of the dark matter constituents have decoupled from the
background RW expansion due to clustering in the process of structure
formation in the universe. Thus we set $t_{\mathrm end} = t_{\mathrm galaxy}$, 
(where $t_{\mathrm galaxy}$ is the time scale of galaxy formation\cite{books}).
Note that the process of wave function collapse
does not of course end at this time, but since the dark matter constituents
are not homogeneously distributed any longer, the rate given by 
Eq.(\ref{darkm1}) ceases to be valid beyond this epoch.
Thus, setting $t_i = t_{\mathrm EW}$, and 
$t_{\mathrm end} = t_{\mathrm galaxy}$, 
and upon performing the integral in Eq.(\ref{darkm2}) 
(from $t_{\mathrm EW} to t_{\mathrm EQ}$ first using $R \propto t^{1/2}$ in the
radiation dominated era, and then, with matching boundary
conditions, from $t_{\mathrm EQ}$ to $t_{\mathrm galaxy}$ 
using $R \propto t^{2/3}$ in the matter dominated era), 
we obtain
\be
\rho_m \simeq (\rho_m)_{\mathrm galaxy} + \frac{\b t_{\mathrm EQ}}{R^3_{\mathrm galaxy}}
\label{darkm3}
\ee
where $(\rho_m)_{\mathrm galaxy}$ is the observed density of dark matter
at the era of galaxy formation, and $t_{\mathrm EQ}$ is the epoch of
matter-radiation equality. Note that
the expression for $\b$ given in Eq.(\ref{enrate}) contains
$M$ that is the total mass of all dark matter in the universe.
Making use of the fact that dark matter contributes to about $50$\% of
the critical density around the era of galaxy formation, we set 
$M/R^3_{\mathrm galaxy} = 0.5 \rho_c$ in Eq.(\ref{darkm3}). 
Substituting the standard value\cite{books} of $t_{\mathrm EQ}=10^{11}s$, 
one gets
\be 
\rho_m \approx 0.5 \rho_c + {\mathcal O}(10^{-22}) \rho_c
\label{darkm4}
\ee
where $\rho_c$ is the critical energy density at $t_{\mathrm galaxy}$.
One sees that the increase of the matter energy density by the 
dynamical collapse process is negligible,  
 which retains its standard value (of the
order of $0.5 \rho_c$) at this era. 

At this stage it may be noted that through our Eqs.(\ref{darkm1}--\ref{darkm4}),
we have calculated the net effect of the wave function collapse process on
the dark matter energy density at the epoch of galaxy formation in the
universe. We have seen that the dark matter energy does not increase in any
non-negligible way. However, conservation of the energy liberated by the 
scalar field aids in the expansion of the universe, as we show now. In order 
to compute the total energy liberated by the
scalar field $\phi$ from the era $t_{\mathrm EW}$ to the era $t_{\mathrm galaxy}$,
we now integrate its instantaneous rate of energy liberation (the scalar
field pumps in energy at this rate throughout the above span of time,
regardless the state of the individual constituents of dark matter particles) 
obtained from Eq.(\ref{conserv}), i.e.,
\be
\dot{\rho}_{\phi} = -\frac{\b}{R^3}
\label{darken1}
\ee  
during the time from 
$t_i = t_{\mathrm EW}$ to $t_{\mathrm end} = t_{\mathrm galaxy}$.  Again using
the relations $R \propto t^{1/2}$, and $R \propto t^{2/3}$, in the radiation
and matter dominated eras, respectively, and using
Eqs.(\ref{enrate}) and (\ref{conserv}), we obtain the total magnitude of the 
energy liberated by the field $\phi$ till the era of galaxy formation to be
\ben
(\rho_{*})_{\mathrm galaxy} \equiv \int_{t_{\mathrm EW}}^{t_{\mathrm galaxy}} 
(-\dot{\rho}_{\phi}dt) \\\nonumber
\qquad \simeq \frac{2\b t_{\mathrm EQ}}{R^3_{\mathrm galaxy}}\biggl(\frac{t_{\mathrm galaxy}}
{t_{\mathrm EQ}}\biggr)^2 \biggl(\frac{t_{\mathrm EQ}}{t_{\mathrm EW}}
\biggr)^{1/2} 
\label{darken2}
\een
Putting in the values of $t_{\mathrm galaxy}= 10^{16}s$, and
$t_{\mathrm EW} = 10^{-10}s$\cite{books}, one obtains
\be
(\rho_{*})_{\mathrm galaxy} \approx  (\rho_m)_{\mathrm galaxy}
\label{darken3}
\ee
where the right hand side denotes the standard amount of dark matter (obtained
from usual scaling due to expansion of the universe) at this epoch. Therefore, 
the excess energy $\rho_{*}$ forms a significant part ($\sim 50\%$)
of the total energy density at about the era of galaxy formation. 
(Note that galaxy formation is a continuous process, but we have set 
$t_{\mathrm end} = t_{\mathrm galaxy} \approx 10^{16}s$ implying a cut-off
beyond which more than half of the total matter density in the universe
is gravitationally clustered, and hence the rate of energy gain
given by Eq.(\ref{darkm1}), as also the rate of energy liberation given by
Eq.(\ref{darken1}) would no longer be valid.)
This energy
liberated by the scalar field does not, of course, add to the matter energy, 
(as we have seen from Eq.(\ref{darkm4}), but nonetheless does contribute to 
the expansion of the universe. 

If the RW expansion takes place in the standard adiabatic 
manner\cite{books}, the
total energy density $\rho_T$ of all the constituents and the pressure $p$ 
must 
satisfy the relation
\be
\frac{d}{dt}(\rho_T R^3) = - p \frac{d}{dt}(R^3)
\label{adiabat}
\ee
Around the time of galaxy formation in the matter 
dominated era, $\rho_T \simeq \rho_m + \rho_{*}$  
(assuming that any remnant energy residing in the scalar field
is negligible compared to $\rho_m$ and $\rho_{*}$) and
$p_m \approx 0$ since the energy liberated through wave function collapse is
unable to substantially increase the kinetic energy or temperature 
of the dark matter constituents (similar to the wave function collapse of
ordinary matter as verified by the results of laboratory 
experiments\cite{exptcsl}). Using
Eqs.(\ref{darkm1}) and (\ref{darken1}) in Eq.(\ref{adiabat}) one obtains,
\be
p = p_{*} = - \rho_{*} - \frac{2\b}{3R^2 \dot{R}}
\label{eqstate}
\ee
Since the second term in the Eq.(\ref{eqstate}) falls off as $1/t$, it follows
that the ($w = p/\rho$) parameter approaches 
($w = -1$) asymptotically.  The equation of state for the ``dark''
energy (DE) around the era of galaxy formation ($t_{\mathrm galaxy}$) is hence 
given by
\be
p = p_{*} = - \rho_{*} - \frac{\b t_{\mathrm galaxy}}{R^3_{\mathrm galaxy}}
\simeq - \rho_{*} - {\mathcal O}(10^{-17})\rho_{*}
\label{eqstate2}
\ee  
resembling closely the equation of state for the
cosmological constant\cite{books}. Till the time 
Eq.(\ref{darken1}) is approximately valid, the DE density increases
as $\rho_{*} \sim -1/t$ in the matter
dominated era. Hence, the liberated dark
energy $\rho_{*}$ with Eq.(\ref{eqstate2}) as its equation of state
can generate the accelerated expansion of the universe once it exceeds the
dark matter density around the era of galaxy formation.

\section{Observational implications}

Beyond $t_{\mathrm galaxy}$, the
matter energy density $\rho_m$ falls off as $1/t^2$, whereas
the dynamical collapse process for the matter continues adding to the ``dark''
energy (DE), albeit with a 
different rate from that given in Eq.(\ref{darken1}). The computation of
such a rate would need to take into account the back-reaction\cite{backreac} 
of structure
formation on the Robertson-Walker metric, which is beyond the
scope of our present analysis.
Assuming a $\L$CDM model of the universe, the 
time $t_{\mathrm galaxy}(= 10^{16}s)$ corresponds 
to a red-shift $z_{\mathrm galaxy} \approx 13$. The equation governing the 
evolution of the DE density $\rho_{*}(z)$ (from Eq.(\ref{darken1})) is  
\be
\label{DE-z}
{d\,\rho_{*} \over d\,z}=-{\b\,(1+z)^2\over H(z)}
\ee
where $H(z)$, the Hubble parameter at red-shift z is  
\be
H(z)=H_0\sqrt{\O_m(0) (1+z)^3 + {\rho_{*}(z)\over \rho_c(0)} },
\ee
$H_0,\,\O_m(0)$ and $\rho_c(0)$ being the Hubble parameter, the 
non-relativistic matter density fraction and the critical density
at the current time ($z=0$). A numerical integration of Eq.(\ref{DE-z}) 
starting from $z=z_{\mathrm galaxy}$ when $\rho_{*}(z)$ 
was half of the critical density at $z_{\mathrm galaxy}$  to  $z=0$ does not 
lead to the requisite amount of DE ($0.73\,\rho_c(0)$) at the present time. 
This just shows that a naive extrapolation of the rate equation 
(Eq.\ref{darken1}) beyond $t_{\mathrm galaxy}(= 10^{16}s)$ is not useful
as the dark matter particles will decouple from the FRW expansion.

Recent additions to SN Ia data sets obtained from various projects like 
the Hubble Space Telescope (HST) ($z\geq1$) \cite{hst}, the Supernova Legacy 
Survey (SNLS) ($z\leq1$) \cite{snls} and the ESSENCE SN Ia survey ($z\leq0.7$) 
\cite{essence} place strong  constraints on the DE equation-of-state 
index $w(z)$,  its variation with red-shift and also the epoch of transition 
from the matter-dominated decelaration phase to the negative-pressure 
DE-dominated acceleration phase. Let us now 
obtain some  order-of-magnitude estimates of the variation of 
$w(z)(=p_{*}(z)/\rho_{*}(z))$ with redshift and also the red-shift of 
transition to the accelerating phase determined by the cross-over of the 
deceleration parameter $q(z)$ from positive to negative values. To this end, 
we first model the DE density evolution by a linear law as follows. 
Let $\O_{*} (z)$ be the ratio of the DE density at any $z$ with respect to 
the current value of the critical density. Then,
\be
\label{Omega-z}
\O_{*} (z)\equiv{\rho_{*}(z)\over \rho_c(0)}=A_0 +A_1\,(1+z),
\ee
where the constants $A_0\,\&\,A_1$ are determined by the two 
conditions: $\O_{*} (0)=0.73$ and the equality of the dark-matter and 
DE at $z_{\mathrm galaxy}$, {\textit i.e.}
\be
\O_{*}(z_{\mathrm galaxy})={1\over2}\left[\O_m (1+z)^3+ \O_{*} (z)\right]_{z=z_{\mathrm galaxy}}
\ee 
Using Eq.(\ref{Omega-z}) in the DE equation-of-state (easily obtained from 
Eq.(\ref{eqstate}) we find that ${d\,w\over d\,z}$ is  ${\cal O}(10^{-17})$ 
for all $z$ upto $z_{\mathrm galaxy}$ which is in conformity with (HST) 
observations which rule out rapidly evolving DE at early times ~\cite{hst}. 
The epoch of transition from deceleration to acceleration $z_{T}$ is defined by 
\be
q({\mathrm z_{T}})\equiv {1\over2}+{4\,\pi \over H^{2}(z_{T})}\,p_{*}(z_{T})=0,
\ee
which, using Eq.(\ref{Omega-z}), comes out to be $z_{T} \approx 19$. It is 
interesting that the above  prediction of our model
 from this crude estimate is approximately within an order of magnitude of 
the $\L$CDM value of  $z_{T}=0.73$ and observational estimates from joint 
analysis of  SN Ia and CMB data \cite{Alam04}: $z_T=0.39\pm 0.03$ for the 
best-fit of the DE models considered  and $z_T=0.57\pm 0.07$ on 
assuming $\L$CDM priors on the $\O_m(0)\, \& \,H_0$ at the present epoch.
As stated earlier, more refined calculations incorporating the effects
of back-reaction\cite{backreac} of structure formation on the associated 
energy liberation rate are required to obtain accurately the evolution
of $\O_{*} (z)$ at low red-shifts in order to confront our model with
present observations.

\section{Conclusions}

To summarize, the above calculations show that dynamical wave function 
collapse\cite{csl,masscsl,rcsl} when applied to the constituents of dark matter
in the universe,
offers a possiblity for the generation of dark energy responsible for the
present acceleration of the universe\cite{accel}.  
A unified framework for dark
energy and dark matter has been presented since in this
approach the former is generated through the interaction
of the latter with a cosmic scalar field. This scheme, though resembling
in spirit some other unified approaches to dark matter and dark 
energy\cite{chaplygin}, is formally quite distinct from them.
Also, unlike in many other models of
dark energy involving the scalar field\cite{DErev,quint,kessence}, 
construction of complicated
potentials is avoided, and the requisite magnitude of dark energy with equation
of state ($w = -1$)  emerges at about the era of galaxy formation.
This energy which could reside either as a kinetic or a potential energy 
component of the scalar field, is liberated  
in a scheme of quantum mechanical disentanglement of the constituents of
dark matter. The scheme presented here combines some essential features of
two hitherto distinct solutions offered respectively for the quantum 
measurement problem\cite{measure} and the dark energy problem\cite{accel}.

Of course, more detailed
calculations are needed to develop our model further. 
For this purpose it
should be particularly useful to evaluate the energy liberated after
the era of galaxy formation up to the present time. 
Such calculations will have to take into account the complex technical
and conceptual aspects\cite{hartle} of the interface of quantum coherence and 
gravitational collapse. Further, the full dynamics of the scalar field
including its possible interactions with matter has to be considered in the 
setting of the expanding universe in order to make our analysis more
comprehensive. Recently, the scheme of dynamical localization 
through a quantized scalar field has been developed \cite{qcsl} which
promises to have rich consequences on gravitational physics and the
physics of the early universe. It might be worthwhile to further develop
our idea in the context of such a scheme. 
The application of quantum entanglement in the dark energy problem has
also been considered in other different contexts\cite{raha}.

Finally, we wish to emphasize that the transfer of energy between
the ``environment'' and the ``system'' is a generic feature of the
quantum
decoherence paradigm\cite{decohenerg}. Though it may
be coincidental for our present simplistic calculation
to yield the requisite magnitude of dark energy, it is relevant to note that
the mathematical structure of the collapse models\cite{csl,masscsl,rcsl}
has striking
similarities\cite{simil} (in spite of interpretational differences) with
that of other decoherence schemes such as the quantum 
state demolition  approach\cite{qsd}, 
and the consistent histories approach\cite{consistent}.
Of course, detailed microscopic modelling of the ``system-environment''
interaction processes in the cosmological background are needed in the
context of the various models\cite{qsd,consistent,rcsl} to confront the
idea of decoherence
induced dark energy with observational data\cite{cmbr-lss,DErev}.
Our present analysis aims
to show that quantum wave function collapse may indeed play a
role in the emergence of the accelerating phase of the universe.

{\it Acknowledgements}: We would like to acknowledge support from the DST
project SR/S2/PU-16/2007.

 
\end{document}